\documentclass[12pt,epsfig]{article}
\newcommand{\ba}{\begin{array}{c}}
\newcommand{\baz}{\begin{array}{cc}}
\newcommand{\bad}{\begin{array}{ccc}}
\newcommand{\bav}{\begin{array}{cccc}}
\newcommand{\ea}{\end{array}}

\usepackage{amssymb} 
\usepackage{epsfig}
\usepackage{float}
\newcommand{\be}{\begin{equation}}
\newcommand{\ee}{\end{equation}}
\newcommand{\bea}{\begin{eqnarray}}
\newcommand{\eea}{\end{eqnarray}}
\begin{document}

\begin{center}
\bf {Recursive parameterisation and invariant phases 
of unitary matrices }
\end{center}

\begin{center}
C. Jarlskog 
\end{center}

\begin{center}
{\em Division of Mathematical Physics\\
LTH, Lund University\\
Box 118, S-22100 Lund, Sweden}
\end{center}

\begin{abstract}

We present further properties of a previously proposed 
recursive scheme for parameterisation of 
n-by-n unitary matrices.
We show that the factors in the recursive formula
may be introduced in any desired order. The method is used
to study the invariant phases of unitary matrices. 
The case of four-by-four unitary matrices is investigated
in detail. We also address the question of how to
construct symmetric unitary matrices using
the recursive approach.

\end{abstract}

\section{Introduction}

Unitary matrices play a central
role in physics. For example, 
the Standard Model of particle physics 
is defined by a $SU(3) \times SU(2) \times U(1)$ symmetry
group and many popular grand unified models
are again based on unitary symmetries. Indeed explicit
representations of unitary matrices are often so badly needed
that there is already a vast literature on the subject
(see, for example \cite{books} and references therein).

Recently we have presented a simple-looking recursive 
parameterisation of general
n-by-n unitary matrices \cite{ceja05}, applicable also,
of course, to subcategories 
such as special unitary matrices and orthogonal
matrices which are of great importance in physics.

In recent publications Fujii and his collaborators
\cite{fujii} have found that the parameterisation in
\cite{ceja05} looks interesting for constructing
unitary gates for quantum computation
but for that purpose more study is needed.
In this paper we present further results on
the structure of the recursive parameterisation 
hoping that it will be useful for future
applications. You use the method to study
the "invariant phases" (to be defined below) 
of unitary matrices, by considering the symmetries of
the recursive parameterisation. Subsequently, we give
detailed attention to the case of four-by-four unitary
matrices.

It should be emphasised that {\it all} 
parameterisation of a general n-by-n
unitary matrix are equivalent to each another. However, for a
specific application a certain parameterisation 
may be more convenient
than others. Therefore, it is important to
provide new parameterisations, a topic which has been
addressed by other authors as well. See, for example, \cite{dita},
a paper which contains an extended list of references
and presents yet another representation 
of unitary matrices.

\section{The parameterisation}   

A general n-by-n unitary matrix $X^{(n)}$ may be expressed 
as a product of three unitary matrices,
\begin{equation}
X^{(n)} = \Phi^{(n)}(\vec{\alpha}) V^{(n)} \Phi^{(n)}(\vec{\beta})
\label{defx}
\end{equation}
where the matrices $\Phi$ are diagonal unitary matrices, 
\begin{equation} 
\Phi^{(n)}(\vec{\alpha}) = diag  
(e^{i\alpha_1}, e^{i\alpha_2}, ..,  e^{i\alpha_n})
\label{defphi}
\end{equation}
$\Phi(\vec{\beta})$ is defined analogously, the $\alpha$'s and
$\beta$'s being real. We shall refer to $\Phi$'s as
external (pure phase) matrices. 

The matrix $X^{(n)}$ has $n^2$ real
parameters. The quantities $\vec{\alpha}$
and $\vec{\beta}$ take care of $2n-1$ of these parameters 
because only the sums $\alpha_i+ \beta_j$ 
enter, where $i$ and $j$ run from $1$ to $n$. The remaining
$(n-1)^2$ real parameters reside in the non-trivial 
matrix $V^{(n)}$ which was the 
subject of the study presented in \cite{ceja05} and
will be further investigated in this paper. For simplicity,
whenever no confusion may arise, we refer 
to $V^{(n)}$ as the most general
n-by-n unitary matrix leaving out the qualifying statement
that this is only true modulus the external matrices
$\Phi(\vec{\alpha})$ and $\Phi(\vec{\beta})$. In \cite{ceja05}
it was shown that the matrix $V^{(n)}$ may be 
written in the form \footnote[1]{Note that the 
notations in this paper are
a simplified version of that in \cite{ceja05}}
\begin{equation}
V^{(n)}= A_{n,2}A_{n,3}... A_{n,n-1}A_{n,n}
\label{vn}
\end{equation}
where the $A_{n,k}$ are unitary matrices defined by 
\begin{equation}
A_{n,k} = \left( \begin{array}{cc} {\mathbb A}^{(k)} & 0 \\
0 & I_{n-k} \end{array}\right) 
\end{equation}
Here $I_{n-k}$ is the unit matrix of order $n-k$. For
$k=n$ this unit matrix is absent.
${\mathbb A}^{(k)}$ is a k-by-k unitary matrix
\begin{equation}
{\mathbb A}^{(k)} \equiv
\left( \begin{array}{cc} 1-(1-c_k)\vert A^{(k)}><A^{(k)} 
\vert & ~s_k \vert A^{(k)}> \\
-s_k < A^{(k)} \vert & ~c_k 
\end{array} \right)
\label{defmatha}
\end{equation}
Here $c_k$ and $s_k$ stand for cosine and sine of an
angle denoted by $\theta_k$. Furthermore, $\vert A^{(k)}>$ is a $k-1$
dimensional complex vector normalised to one,
\begin{equation}
\vert A^{k)}> = \left( \begin{array}{c}
a_1^{(k)} \\ a_2^{(k)} \\ .\\.\\ a_{k-1}^{(k)} \end{array}
\right), ~~~~ < A^{k)} \vert A^{k)}> =1
\label{defa}
\end{equation}
and $(\vert A^{(k)}><A^{(k)}\vert)_{ij} = a_i^{(k)}a_j^{(k)\star}$.
We shall refer to $\vert A^{(k)}>$ as the characteristic
vector of order $k$.
 
The parameter counting was presented in \cite{ceja05}
where it was shown that $V^{(n)}$, thus obtained, is
the most general n-by-n unitary matrix, again modulus
the external matrices $\Phi$. The essential point is
that $\vert A^{k)}>$ introduces $2(k-2)$ real parameters
and not $2(k-1)$, the reason being that
it is normalised and its overall phase can
be absorbed into the definition of the external matrices
$\Phi$.

To summarise, in this recursive parameterisation the
n-by-n unitary matrix is represented
by a product of n-1 unitary matrices, each with
its own angle $\theta$ and characteristic vector
$\vert A>$ while, for example, in the
conventional approach in particle physics one
would write the matrix as a product of at least
$n(n-1)/2$ matrices, these being Euler rotation
matrices, (n-1)(n-2)/2 of them modified by 
phases (see, for example \cite{pdg}
and references therein). Note also that in Ref. 
\cite{dita} the matrix
is parameterised by a product of $n$ diagonal unitary matrices
interlaced with $n-1$ orthogonal matrices. 

\subsection{Reordering of the factors}
\label{reordering}

At the first sight, the recursive parameterisation appears
highly ordered and rigid. In Eq.(\ref{vn}), 
the two-by-two structure 
is immediately followed by the three-by-three and so on.
Actually, we may write these
factors {\it in any order} we wish
by observing that the order of factors in a
given product, $A_{n,r}A_{n,s}$, may be flipped as follows. 
For $r<s$ we have
\begin{equation}
A_{n,r}A_{n,s} =(A_{n,r}A_{n,s}A^\dagger_{n,r})A_{n,r}
\equiv A^\prime_{n,s}A_{n,r} 
\end{equation}
where $A^\prime_{n,s}$ has the same form as
$A_{n,s}$. The two characteristic vectors appearing in these
matrices are related by a unitary rotation,
\begin{eqnarray}
\vert A^{\prime (s)}>& = &\hat{{\mathbb A}}^{(r)}
\vert A^{(s)}> \\
\hat{{\mathbb A}}^{(r)} &=&
\left( \begin{array}{cc} {\mathbb A}^{(r)} & 0 \\
0 & I_{s-1-r} \end{array}\right)
\end{eqnarray}  
For the case $s <r $ we have
\begin{equation}
A_{n,r}A_{n,s} = A_{n,s}(A^\dagger_{n,s}A_{n,r}A_{n,s})
\equiv A_{n,s} A^{\prime \prime}_{n,r}
\end{equation}
where now the two characteristic vectors 
are related by 
\begin{equation}
\vert A^{\prime \prime (r)}> = \hat{{\mathbb A}}^{\dagger(s)}
\vert A^{(r)}> 
\end{equation}
Obviously, by inserting as many factors 
$A^\dagger_{n,j}A_{n,j}=1$ as needed 
in the recursion formula, Eq.(\ref{vn}), one may move the factors
around as one wishes. The upshot is that in the reordering
process a factor of lower rank simply "tunnels" through that of a
higher rank without being affected but induces a unitary rotation
of the characteristic vector of the latter. Thus the ensuing
parameterisation remains the most general one. Note that the
angles $\theta_k$ remain invariant under reordering.

The recursive parameterisation looks highly asymmetric. However,
using the above reordering procedure, one may construct
manifestly symmetric unitary matrices (see Appendix). 
       
\subsection{Further properties of the recursive parameterisation }
\label{properties}

We wish to study, in more detail, the properties 
of the matrices ${\mathbb A}^{(k)}$ in Eq.(\ref{defmatha}) as
these are the building blocks of the recursive parameterisation.

To begin with we simplify the notation, to
avoid indices, and 
introduce a generic matrix ${\mathbb A}$ defined by
\begin{equation}
{\mathbb A} \equiv
\left( \begin{array}{cc} 1-(1-c)\vert A><A 
\vert & ~s \vert A> \\
-s < A \vert & ~c 
\end{array} \right)
\end{equation}
Here, as usual, $c$ and $s$ stand for cosine and sine of an
angle respectively. The angle itself will be denoted by $\theta$. 
Defining
\begin{equation}
Y \equiv \vert A><A \vert
\end{equation}
we have that $Y$ is hermitian and satisfies 
\begin{equation}
Y \vert A> = \vert A >, ~~
Y^2 = Y, ~~trY =1, ~~det Y =0
\label{propy}
\end{equation} 
where the vanishing of the determinant is, of course,
only valid when $Y$ is a matrix and not just a number
as is the case when $\vert A>$ is one dimensional.
Following Fujii \cite{fujii}, we introduce 
a matrix $\mathbb G$ 
which generates ${\mathbb A}$,  
\begin{equation}
{\mathbb G} \equiv \left( \begin{array}{cc}
0 & -i \vert A> \\
i < A \vert & 0
\end{array} \right)
\end{equation}
This matrix is hermitian and satisfies
${\mathbb G}^3 = {\mathbb G}$. A simple
computation, using Eqs.(\ref{propy}), yields
\begin{equation}
{\mathbb A} = e^{i\theta {\mathbb G}} = 
1 + is {\mathbb G} - (1-c) {\mathbb G}^2
\label{gexpan}
\end{equation}
This relation is reminiscent of the expansion
of exponentials containing Pauli matrices $\sigma$ 
(in a short-hand notation, 
$e^{i\theta \sigma} = c + is \sigma $). The
essential point here is that the series expansion 
of $e^{i\theta {\mathbb G}}$, for 
arbitrarily ${\mathbb G}$,  
terminates rapidly and does
not continue for ever as the exponentials often tend to do.
Furthermore we have 
\begin{equation}
tr {\mathbb G} =0, ~~ tr {\mathbb G}^2 =2, ~~det {\mathbb A} =1
\end{equation}
Note also that, for a fixed ${\mathbb G}$, the matrix
${\mathbb A}$ is Abelian with respect to $\theta$,
\begin{equation}
{\mathbb A}(\theta_i) {\mathbb A}(\theta_j) = 
{\mathbb A}(\theta_i + \theta_j) 
\end{equation}
and
\begin{equation}
 {\mathbb A}^{-1}(\theta) = {\mathbb A}(-\theta)
\end{equation} 
To rewrite the 
recursion formula, Eq.(\ref{vn}), in terms
of ${\mathbb G}$ and ${\mathbb A}$ we
must attach appropriate indices to our generic 
$\mathbb G$ (or $\mathbb A$) to distinguish the 
relevant factors. We
introduce
\begin{equation}
{\mathbb G}_{n,k} =
\left( \begin{array}{cc}  \left(
\begin{array}{cc}
 0 & -i \vert A^{(k)}>  \\
i < A^{(k)} \vert & 0
\end{array}  \right) 
& 0 \\ 0 & 0\end{array}
\right)
\label{gnk}
\end{equation}
where the required number
of zeros have been added to make ${\mathbb G}_{n,k}$ 
an n-by-n matrix.
This yields that the factor ${A}_{n,k}$ in the recursion
formula Eq.(\ref{vn}) is given by
\begin{equation}
{A}_{n,k} \equiv
e^{i\theta_k {\mathbb G}_{n,k}} = 
1 + is_k  {\mathbb G}_{n,k} - (1-c_k)~
{\mathbb G}_{n,k}^2  
\label{gexpanp}
\end{equation}

\section{Invariant phases of unitary matrices}

The invariant phases of a unitary n-by-n matrix $V^{n}$ are
defined as those phases of the matrix that cannot be "removed" 
with any choice of the external phase matrices 
$\Phi$ in Eq.(\ref{defx}). These phases play an important role
in particle physics as they are measurable quantities
related to CP violation (for a review see, for example,
Ref. \cite{cpboken}). 

Given a unitary matrix, the simplest way to detect 
the presence of invariant phases in it is to construct
\begin{equation}
(\alpha \beta; jk) \equiv Im (V_{\alpha j}V_{\beta k} V^\star_{\alpha k}
V^\star_{\beta j} )
\label{defjs}
\end{equation}
where we have suppressed the superscript $n$. The
symbols $\alpha,\beta$ and $j,k$ now refer to rows and columns
of the matrix and the indices are not summed.
These imaginary parts are 
manifestly invariant under multiplication by the external
phase matrices. Therefore if any of them is nonzero that
would be a signal of the presence of a nonremovable phase
in the matrix.
We refer to these imaginary parts as invariant phases
of the matrix
instead of calling them invariants of the matrix
that contain nonremovable phases. One may easily construct higher
order invariants, containing properly chosen, six or more, elements
of the matrix but these are in general reducible to the above
set unless the matrix would have vanishing 
elements. For example, for  $ V_{\beta j} \ne 0$ 
\begin{eqnarray}
Im (V_{\alpha j}V_{\beta k} V_{\gamma l} 
V^\star_{\alpha k} V^\star_{\beta l} V^\star_{\gamma j} )
&=& {1 \over \vert V_{\beta j} \vert^2}
Im \left\{ (V_{\alpha j}V_{\beta k} V^\star_{\alpha k} V^\star_{\beta j})
(V_{\beta j} V_{\gamma l} 
V^\star_{\beta l} V^\star_{\gamma j})\right\} \\
&=& {1 \over \vert V_{\beta j} \vert^2}
\left\{ (\alpha \beta, jk) \langle \beta \gamma, jl \rangle +
\langle \alpha \beta, jk \rangle (\beta \gamma, jl)\right\} 
\end{eqnarray}
where none of the indices is summed and
\begin{equation}
\langle \alpha \beta; jk \rangle \equiv  
Re [V_{\alpha j}V_{\beta k} V^\star_{\alpha k}
V^\star_{\beta j} ]
\end{equation}
These real parts are also invariant under the action
of the external matrices.
The above reduction would not work if $V_{\alpha j}=0$
but then the analysis is much simpler to begin with
(see below) as the matrix contains fewer invariant
phases.

Returning to the simplest invariant phases,
there are altogether $[n(n-1)/2]^2$
quantities $(\alpha \beta; jk)$, because these are antisymmetric 
under the interchange of the row indices,  
$\alpha \leftrightarrow \beta$,
as well as under the interchange of the column
indices, $j \leftrightarrow k$.
However, we know that the most general
$V^{(n)}$ has "only" $(n-1)(n-2))/2$ independent invariant
phases. One may therefore look for
$(n-1)(n-2))/2$ independent $(\alpha \beta; jk)$'s and
use them as a basis for expressing the remaining ones.

As mentioned above, the invariants in Eq.(\ref{defjs}) 
play an essential role in the
$n$-family version of the
Standard Model of particle physics as they are 
measurable quantities related to CP-violation.
For the case of $n=3$ there is only one such
quantity 
\begin{equation}
(\alpha \beta; jk) \equiv J \sum_{\gamma, i}^{}
\epsilon_{\gamma \alpha \beta} \epsilon_{ijk}
\end{equation}
The row and column unitarity conditions
for a three-by-three unitary matrix
define six triangles. One may show that \cite{stora}
all these triangles have the same area and this
unique area equals $J/2$. 
  
For $n=4$ there are 36 possible
invariants $(\alpha \beta; jk)$ but only three independent
ones. In \cite{ceja87a} an attempt was made to find
an appropriate basis and carry through the above programme.
The treatment of this issue is much simpler in the
recursive parameterisation, 
as will be shown in the next section.

\subsection{Invariant phases of four-by-four unitary matrices} 

For $n=4$ we have from Eq.(\ref{vn}) 
\begin{equation}
V^{(4)}= A_{4,2}A_{4,3}A_{4,4}
\label{v4}
\end{equation}
where each factor comes with its own $\theta$
and characteristic vector $\vert A> $. We denote the latter by
\begin{equation}
\vert A^{(2)}> =1,~~~
\vert A^{(3)}> = \left( \begin{array}{c}
x_1 \\ x_2 \end{array}
\right),~~~
 \vert A^{(4)}> = \left( \begin{array}{c}
y_1 \\ y_2 \\ y_3 \end{array}
\right)  
\end{equation}
remembering that $x$'s and $y$'s are complex
numbers and $< A^{(k)} \vert A^{(k)}> =1$, $k=2,3,4$.
We shall now spell out this four-by-four matrix in
order to exhibit its symmetries in a manifest fashion.
This will also enable us to understand the general case
of n-by-n matrices.  
Eq.(\ref{vn}) yields 
\begin{eqnarray}
V^{(4)} & = & \left( \begin{array}{cccc}
c_2 & s_2 & 0 & 0 \\
-s_2 & c_2 & 0 & 0 \\
0 & 0 & 1 & 0 \\
0 & 0 & 0 & 1
\end{array}  \right)
\left( \begin{array}{cccc}
1-(1-c_3) x_1 x^\star_1 & -(1-c_3) x_1 x^\star_2 & s_3 x_1 & 0 \\
-(1-c_3) x_2 x^\star_1 & 1-(1-c_3) x_2 x^\star_2 & s_3 x_2 & 0 \\
-s_3 x^\star_1 & -s_3 x^\star_2 & c_3 & 0 \\
0 & 0 & 0 & 1
\end{array}  \right) \nonumber\\
& \times & 
\left( \begin{array}{cccc}
1-(1-c_4) y_1 y^\star_1 & -(1-c_4) y_1 y^\star_2 
& -(1-c_4) y_1 y^\star_3 & s_4 y_1 \\
-(1-c_4) y_2 y^\star_1 & 1-(1-c_4) y_2 y^\star_2 
& -(1-c_4) y_2 y^\star_3 & s_4 y_2 \\
-(1-c_4) y_3 y^\star_1 & -(1-c_4) y_3 y^\star_2
& 1-(1-c_4) y_3 y^\star_3 & s_4 y_3  \\
-s_4 y^\star_1 & -s_4 y^\star_2 & -s_4 y^\star_3 & c_4 
\end{array}  \right)
\label{v4p}
\end{eqnarray}
We now focus on the symmetries of this matrix.
By symmetries we mean transformations
that leave $V^{(4)}$ invariant modulus the external 
matrices $\Phi$ in Eq.(\ref{defx}). A simple inspection
shows that this
matrix has two such symmetries, denoted by $S_1$
and $S_2$ and defined by
\begin{eqnarray}
S_1&:&~~~ \left( \begin{array}{c}
x_1 \\ x_2 \end{array}
\right) ~\rightarrow ~ e^{i\phi_2} \left( \begin{array}{c}
x_1 \\ x_2 \end{array}
\right); ~~y_3 ~\rightarrow ~ e^{-i\phi_2} y_3 \\
S_2 &:&~~~ 
\left( \begin{array}{c} y_1 \\ y_2 \\ y_3 
\end{array} \right) ~
\rightarrow  ~ e^{i\phi_3} \left( \begin{array}{c} y_1 \\ y_2 \\ y_3 
\end{array} \right)
\label{yphase}
\end{eqnarray}
where $\phi_2$ and $\phi_3$ are arbitrary
phases. The indices are to remind us of the dimension
of the corresponding vector. Therefore
the three independent phases in
$V^{(4)}$ can be chosen to be:
\begin{eqnarray}
\omega_1 &= &\phi (x_2) - \phi (x_1) \nonumber \\
\omega_2 &= &\phi (y_2) - \phi (y_1) \nonumber \\
\omega_3 &= &\phi (x_2) + \phi (y_3) - \phi (y_2)
\label{defomega} 
\end{eqnarray}
Here $\phi (x_j)$ and $\phi (y_k)$ denote the phases
of the corresponding parameters. These phases are not
invariant under the above symmetries and thus can't
appear as independent entities in computations of
invariants of $V^{(4)}$ . Allowed to appear are
the $\omega$'s
or any combination of them because these are invariant under
the action of both $S_1$ and $S_2$. 
We may, if we so wish, use the symmetry $S_1$
to rotate the phase of
$x_2$ to zero, and then employ the symmetry $S_2$ to do
the same with the ensuing $y_3$ whereby
$x_2$ and $y_3$ may be taken to be real and say
positive. The invariant phases in this "frame"
are then $\phi (x_1)$, $\phi (y_1)$ and $\phi (y_2)$.
These constitute the maximum number of independent phases that
$V^{(4)}$ can possesses. By imposing further relations on
the angles or the $x$'s and $y$'s this number
could be smaller as shall be considered further below. 

\subsection{Generalisation to larger $n$}

Going one order higher to $n=5$, we have to
introduce the relevant characteristic vector
\begin{equation}
\vert A^{(5)} > \equiv
\left( \begin{array}{c} z_1 \\ z_2 \\ z_3 \\ z_4
\end{array} \right) 
\label{zphase}
\end{equation}
The corresponding matrix $V^{(5)}$ has three symmetries given by
\begin{eqnarray}
S_1 &:&~~~ \left( \begin{array}{c}
x_1 \\ x_2 \end{array}
\right) ~\rightarrow ~ e^{i\phi_2} \left( \begin{array}{c}
x_1 \\ x_2 \end{array}
\right); ~~y_3 ~\rightarrow ~ e^{-i\phi_2} y_3, ~
z_3 ~\rightarrow ~ e^{-i\phi_2} z_3 \\
S_2 &:&~~~ 
\left( \begin{array}{c} y_1 \\ y_2 \\ y_3 
\end{array} \right) ~
\rightarrow  ~ e^{i\phi_3} \left( \begin{array}{c} y_1 \\ y_2 \\ y_3 
\end{array} \right), ~~z_4 \rightarrow e^{-i\phi_3} \\ 
S_3 &:&
\left( \begin{array}{c} z_1 \\ z_2 \\ z_3 \\ z_4
\end{array} \right) ~
\rightarrow  ~ e^{i\phi_4} \left( \begin{array}{c}
z_1 \\ z_2 \\ z_3 \\ z_4 
\end{array} \right)
\label{5families}
\end{eqnarray}
In this case there are six invariant phases. These may be
chosen as
\begin{eqnarray}
\omega_1 &= &\phi (x_2) - \phi (x_1) \nonumber \\
\omega_2 &= &\phi (y_2) - \phi (y_1) \nonumber \\
\omega_3 &= &\phi (z_2) - \phi (z_1) \nonumber \\
\omega_4 &= &\phi (x_2) + \phi (y_3) - \phi (y_2) \nonumber \\
\omega_5 &= &\phi (x_2) + \phi (z_3) - \phi (z_2) \nonumber \\
\omega_6 &= &\phi (y_3) + \phi (z_4) - \phi (z_3)  
\end{eqnarray}
As before, we may use $S_1$ to remove the phase of $x_2$
followed by $S_2$ and $S_3$ to rotate the ensuing $y_3$ and
$z_4$ to be real and say positive. The invariant phases are then
the phases of $x_1$, $y_1$, $y_2$,
$z_1$, $z_2$ and $z_3$. Note that, from the very beginning
we chose the angle $\theta_2$ not to be accompanied by
a phase, i.e., $\vert A^{(2)}> =1$. One can, of course,
leave the phase of $\vert A^{(2)}> $ arbitrary. This will
introduce an extra symmetry which we have not bothered
to write down as it is trivial.

The above procedure may be generalised to arbitrary
order $n$. Without loss of generality,
we may take, for example, the last component of all the
characteristic vectors $\vert A^{(k)} >$ to be real. The invariant
phases are then the phases of the remaining components.
For an n-by-n matrix there are then $1+2+..+(n-2)
= (n-1)(n-2)/2$ such independent phases as expected.

\subsection{The "panel" approach to invariant phases}
\label{panel}

Another approach to constructing the invariant phases
of a matrix is to consider the latter as a lattice.
For the case of $n=4$, considered from now on,
the matrix can be visualised as shown 
\begin{equation}
\begin{array}{ccccccc}
\bullet &  & \bullet & & \bullet & & \bullet \\
& P_{11} & & P_{12} & & P_{13}  &\\
\bullet &  & \bullet & & \bullet & & \bullet \\
& P_{21} & & P_{22} & & P_{23}  &\\
\bullet &  & \bullet & & \bullet & & \bullet \\
& P_{31} & & P_{32} & & P_{33}  &\\
\bullet &  & \bullet & & \bullet & & \bullet
\end{array}
\end{equation}
The bullets denote the sites where the matrix
elements are situated. For example the bullets on
the first row stand for $V_{11}$, $V_{12}$, $V_{13}$ and
$V_{14}$ and so on. The $P$'s denote minipanels
of the matrix, to be described here below. The
invariants of interest to us are again
\begin{eqnarray}
(\alpha \beta; jk) & \equiv & Im [V_{\alpha j}V_{\beta k} V^\star_{\alpha k}
V^\star_{\beta j} ] \\
\langle \alpha \beta; jk \rangle & \equiv & Re [V_{\alpha j}V_{\beta k} V^\star_{\alpha k}
V^\star_{\beta j} ]  
\end{eqnarray}
As mentioned before,
there are 36 quantities $(\alpha \beta; jk)$
(six possible combinations of $\alpha, \beta$ multiplied by 
as many combinations of $j,k$) and we are looking for
a set of three of them such that all the others can
be expressed as functions of them and the real parts
$\langle \alpha \beta; jk \rangle$. This problem was treated
long time ago \cite{ceja87a} and was found to be
rather involved. Here we provide some simplification.
Nine of these invariants (the nearest neighbours) are
explicitly exhibited on our lattice. Their 
analytic form may easily 
be read off from their locations. For example, 
\begin{eqnarray}
P_{11} &\equiv& V_{11}V_{22}V^\star_{12}V^\star_{21} \nonumber \\
P_{12} &\equiv& V_{12}V_{23}V^\star_{13}V^\star_{22} \nonumber \\
P_{22} &\equiv& V_{22}V_{33}V^\star_{23}V^\star_{32} \nonumber \\
P_{32} &\equiv& V_{32}V_{43}V^\star_{33}V^\star_{42} \nonumber 
\end{eqnarray}
and so on. Furthermore
\begin{equation}
P_{ab} = R_{ab} + i J_{ab}
\end{equation}
where $R$ and $J$ denote the real and imaginary parts
of the corresponding $P$. Thus the imaginary parts, in
the notation employed earlier, are
given by
\begin{equation}
J_{11} = (12;12), ~~~J_{12} = (12;23)~~~ 
J_{22} = (23;23)~~~J_{32} = (34;23)
\end{equation}
and so forth. Similar expressions may be written
down for the real parts.

Suppose that none of the matrix elements vanishes. 
Using unitarity conditions we then find
\begin{eqnarray} 
J_{13} - (1 + {R_{11} \over \vert 
V_{12}V_{22} \vert^2}) J_{12} &=&
{R_{12} \over \vert V_{12}V_{22} \vert^2} J_{11} \nonumber \\ 
J_{13} - (1 + {R_{33} \over
\vert V_{33}V_{34} \vert^2}) J_{23} &=& 
{R_{23} \over \vert V_{33}V_{34} \vert^2} J_{33} \nonumber \\ 
J_{31} - (1 + {R_{11} \over 
\vert V_{21}V_{22} \vert^2}) J_{21} &=&
{R_{21} \over \vert V_{21}V_{22} \vert^2} J_{11} 
\nonumber \\ 
J_{31} - (1 + {R_{33} \over
\vert V_{33}V_{43} \vert^2}) J_{32} &=& 
{R_{32} \over \vert V_{33}V_{43} \vert^2} J_{33} \nonumber \\ 
J_{12} - {R_{22} \over \vert V_{32}V_{33} \vert^2} 
J_{32} &=& 
 (1 + {R_{32} \over 
\vert V_{32}V_{33} \vert^2}) J_{22} \nonumber \\
J_{21} - {R_{22} \over \vert V_{23}V_{33} \vert^2} 
J_{23} &=& 
 (1 + {R_{23} \over 
\vert V_{23}V_{33} \vert^2}) J_{22} 
\end{eqnarray} 
In principle, we may take the set $J_{11}, J_{22}, J_{33}$
to constitute a basis and determine the remaining six
$J$'s in terms of them. These equations are 
rather complicated and need further
thought concerning special cases. For example, 
if the matrix is symmetric we only have three equations but,
of course, also
only three unknown, say $J_{12}, J_{13}$ and $J_{23}$.
We are not allowed to divide by vanishing matrix elements and
so forth. Here below, we shall consider a simple and
yet nontrivial example to demonstrate the technique and
to compare it with the recursive approach which is
much simpler and does not require thinking about
the possible pitfalls. 

\subsection{A simple example}
\label{simple}

For simplicity we consider the case where
two of the elements of the four-by-four matrix
are zero, and where these elements are neither on
the same row nor on the same column. All other elements of the
matrix are assumed to be nonzero. Without loss of
generality we may take the two vanishing elements to be
$V_{14}$ and $V_{41}$ which in particle physics 
would correspond to the case where the mixing of the
first and the fourth families is negligible. 
Our lattice, with its nine minipanels, now
looks as follows
\begin{equation}
\begin{array}{ccccccc}
\bullet &  & \bullet & & \bullet & & \odot \\
& J & & J & & 0  &\\
\bullet &  & \bullet & & \bullet & & \bullet \\
& J & & J+J^\prime & & J^\prime  &\\
\bullet &  & \bullet & & \bullet & & \bullet \\
& 0 & & J^\prime & & J^\prime  &\\
\odot &  & \bullet & & \bullet & & \bullet
\end{array}
\end{equation}
Here the $\odot$'s indicate where the vanishing
matrix elements are situated and we have defined
\begin{equation}
J = J_{11} \equiv (12, 12), ~~~ J^\prime = J_{33} \equiv (34,34)
\label{jjp} 
\end{equation}
A simple computation, using unitarity 
relations, gives the imaginary parts of the minipanels as marked
in the lattice. Computing all the imaginary parts, we find
that 19 of the 36 invariants $(\alpha \beta,jk)$ vanish.
The nonvanishing ones, in addition to $J$ and $J^\prime$
defined in Eq.(\ref{jjp}), are
\begin{eqnarray}
-(12,13)=(12,23)=-(13,12)=(13,13)=
 -(13,23)=(23,12)=-(23,13) &=& J\nonumber \\
-(23,24)=(23,34)=-(24,23)=(24,24)=
-(24,34)=(34,23)=-(34,24)  &=& J^\prime
\end{eqnarray}
and
\begin{equation}
(23,23)= J + J^\prime
\end{equation}
as exhibited in the corresponding panel.
Moreover, we find
\begin{eqnarray}
{J^{\prime 2} \over J^2} &=& \vert 
{V_{24} V_{34} \over V_{21} V_{31}} \vert^2= \vert 
{V_{42} V_{43} \over V_{12} V_{13}} \vert^2 
\nonumber \\
&=&({\vert V_{24}\vert^2 + \vert V_{34}\vert^2 
\over \vert V_{12}\vert^2 + \vert V_{13} \vert^2})^2=
({\vert V_{42}\vert^2 + \vert V_{43}\vert^2 
\over \vert V_{21} \vert^2 + \vert V_{31}\vert^2})^2
\end{eqnarray}
It is amusing to note that the unitarity relations for
the above matrix define eight triangles. Using the
method in \cite{stora} one finds
that four of these have each an area equal to $J/2$ while
the area of the other four is $J^\prime /2$. 

We would now like 
to compute $J$ and $J^\prime$. For this purpose
we turn to the recursive parameterisation.   
It turns out that the calculations are
simpler is we take $V_{34}= V_{43} =0$ instead of the above
choice $V_{14}= V_{41} =0$. The two choices are
equivalent as they are related to one another
by interchanges in rows and columns. This amounts
to a relabeling of the matrix elements which 
obviously can't affect the results.  
After finishing the computations we can simply revert
to the former case by interchanging rows one and three
as well as columns one and three. 

In the recursive
parameterisation, Eq.(\ref{v4p}), the conditions
$V_{34}= V_{43} =0$ give
\begin{equation}
y_3 = x_1 y^\star_1 + x_2 y^\star_2 =0
\end{equation}
These condition tell us that $\vert y_1 \vert = \vert x_2 \vert $,
$\vert y_2 \vert = \vert x_1 \vert $ and that
$x_1$ and $y_1$ are relatively real
in the frame where $x_2$ and $y_2$ are taken to be real.
Therefore, there is only one invariant phase, in this example.
We introduce the lattice again
\begin{equation}
\begin{array}{ccccccc}
\bullet &  & \bullet & & \bullet & & \bullet \\
& \hat{J} & & \hat{J} & & 0  &\\
\bullet &  & \bullet & & \bullet & & \bullet \\
& \hat{J} & & \hat{J} & & 0  &\\
\bullet &  & \bullet & & \bullet & & \odot \\
& 0 & & 0 & & 0  &\\
\bullet &  & \bullet & & \odot & & \bullet
\end{array}
\end{equation}
$\odot$'s indicate where the
vanishing matrix elements are situated. 
Furthermore, we have exhibited the imaginary parts of
the minipanels starting with the definition
\begin{equation}
\hat{J} = (12, 12)
\end{equation}
Taking into account the permutations, we find that
$\hat{J}$ here is identical with our previous $J$ that
we wanted to compute. Using the recursive parameterisation
we find 
\begin{eqnarray}
J &= & c_2 c_3 c_4 s_2 s^2_3  Im (x^\star_1 x_2)
\nonumber \\
J^\prime &= & - c_2 c_3 c_4 s_2 s^2_4 Im (x^\star_1 x_2) 
\end{eqnarray}
Thus
\begin{equation}
{J^\prime \over J} = -{s^2_4\over s^2_3}
\end{equation}
For comparison, note that for a general three-by-three matrix
(which we may obtain from  
Eq.(\ref{v4p}) by putting $\theta_4 =0$) the unique invariant
is given by $J^{(3-fam)} = c_2 c_3 s_2  s^2_3 Im (x^\star_1 x_2)$.

The recursive parameterisation allows us to compute
all the imaginary parts for the most general case,
i.e., irrespectively of whether the matrix has
zeros or not. 
We find, for the general four-by-four matrix,
parameterised as in Eq.(\ref{v4p})
\begin{eqnarray}
(34,34)&=& c_3 c_4 s_3 s^2_4 \vert y_3 \vert 
\left\{ \vert x_2 y_2 \vert sin \omega_3 + 
\vert x_1 y_1 \vert sin (\omega_3+ \omega_2 - \omega_1)\right\}
\nonumber \\
(34,24)&=& c_4 s_3 s^2_4 \vert x_2 y_2 \vert
\left\{ s_3 \vert x_1 y_1 \vert sin(\omega_1 - \omega_2)
-c_3 \vert y_3 \vert sin\omega_3 \right\}
\end{eqnarray}
where the angles $\omega_j$ are as defined in
Eq.(\ref{defomega}). We thus see that, as expected, 
only the invariant phases appear in these relations.
We do not quote the remaining imaginary parts 
$(\alpha \beta, jk)$. The important point is that
all of them are functions of $\vert x_j \vert $,
$\vert y_j \vert $ and the three $\omega$'s, as
expected.

\section{Conclusions}

In this paper, we have presented further properties
of the recursive parameterisation of unitary
matrices proposed in \cite{ceja05}, where the
matrix is written as a product of $n-1$ matrices
each with its own angle $\theta$ and characteristic
vector $\vert A>$. We have found
that the factors in the recursive formula
may be introduced in any desired order.

Encouraged by the convenience of the recursive method,
we have taken a fresh look at the 
issue of invariant phases of unitary matrices.
After having exhibited the symmetries of the
parameterisation, we have shown how 
the invariant phases of n-by-n matrices can be identified. 
Subsequently, we have paid particular 
attention to the case $n=4$ 
and have compared the results 
with those of an earlier approach based
on "panels" of the matrix. 

The recursive parameterisation has some really
nice features because in some cases it allows the "new physics"
to be introduced in a gentle manner through the
last factor in the recursion formula, a topic
which we are currently studying.
 
In an earlier study \cite{cejacab}, we found 
that there is
a parameterisation that allows one to  
introduce, in a simple way, any desired angle of any of the 
so called unitarity
triangles as one of the parameters in the quark mixing
matrix for three families in the Standard Model
of particle physics. The same
parameterisation allowed 
us to choose the expansion parameter in this
matrix to be $\lambda^2$ instead of 
$\lambda$ that one usually uses \cite{pdg}. 
Indeed $\lambda$ is not so small 
($\lambda =0.2$). Therefore, the recursive
parameterisation may be convenient
whenever expansion in the above parameter is
required, for example in model building or for
construction of quark and lepton mass matrices.  
It turned out that the parameterisation found in
\cite{cejacab},
with the above nice features,
is indeed nothing but the order $n=3$ version of
the recursive parameterisation discussed in
this paper and in \cite{ceja05}.

Finally, in the appendix of this paper, 
we deal with the question
of how to construct manifestly
symmetric unitary matrices in the recursive framework.

\section{Appendix: Symmetric unitary matrices}

We write the symmetric unitary matrix in the form
\begin{equation}
X^{(n)sym} = \Phi^{(n)}(\vec{\alpha}) V^{(n)sym} 
\Phi^{(n)}(\vec{\alpha})
\label{defxsym}
\end{equation}
requiring $V^{(n)sym}$ to be symmetric, as indicated by its
superscript, and that
the external matrices be the same (see Eq.(\ref{defx})).
A general symmetric unitary matrix has $n(n+1)/2$ real
parameters. The external matrix $\Phi$ takes care
of $n$ of them. Thus $n(n-1)/2$ real parameters reside in
$V^{(n)sym}$. 

For a general $V^{(n)}$,  we have  
(see Section \ref{properties}) that the factors
in the recursion formula Eq.(\ref{vn}) may be 
written as
\begin{equation}
{A}_{n,k} =
e^{i\theta_k {\mathbb G}_{n,k}}
\end{equation}
where the generating matrix ${\mathbb G}_{n,k}$ is hermitian.
In order to obtain a symmetric ${A}_{n,k}$
we must impose the additional requirement that
the generating matrix be symmetric. This means that
the corresponding characteristic vector is purely imaginary.
For example, for $k=2$ we obtain
\begin{equation}
{V}^{(2)sym} \equiv {A}^{sym}_{n,2} =
\left( \begin{array}{ccc}
c_2 & is_2 & 0  \\
is_2 & c_2 & 0  \\
0 & 0 & I_{n-2}
\end{array} \right)
\label{v2sym}
\end{equation}
and for $k=3$
\begin{equation}
{A}^{sym}_{n,3}
\left( \begin{array}{cccc}
1-(1-c_3) x^2_1 & -(1-c_3) x_1 x_2 & is_3 x_1& 0 \\
-(1-c_3) x_1 x2_1 & 1-(1-c_3) x^2_2  & is_3 x_2 & 0 \\
is_3 x_1 & is_3 x_2 & c_3 & 0 \\
0 & 0 & 0 & I_{n-3}
\end{array}  \right)
\label{A3sym}
\end{equation}
We have put $\vert A> = i \vert x>$, where the $x$'s are real.
Thus, the construction of symmetric factors in the recursion
formula is a trivial task. But, of course, the product
of these factors will not be symmetric. This defect is
easily remedied by invoking the reordering procedure
described in Section \ref{reordering} in which we showed
that the reordering of the factors in the recursion formula
only amounts to a redefinition of the characteristic vectors.
Therefore, we may write 
\begin{equation}
 V^{(n)sym} ={A}^{sym}_{n,2} {A}^{sym}_{n,3} ...{A}^{sym}_{n,n-1}
{A}^{sym}_{n,n}{A}^{sym}_{n,n-1}...{A}^{sym}_{n,3}{A}^{sym}_{n,2}
\label{vnsym}
\end{equation}
$V^{(n)sym}$ thus obtained is manifestly unitary and symmetric.
We must now count the number of its independent parameters.
Each order $k$ introduces $k-1$ 
real parameters, these being the angle $\theta_k$ 
and $k-2$ components of the corresponding
characteristic vector (one component being
redundant because the vector is normalised).
Therefore the total number of parameters in
$V^{(n)sym}$ is 
\begin{equation}
\sum_{k=2}^{n} (k-1) = n(n-1)/2
\end{equation}
as expected. Adding into this number the $n$ parameters 
coming from
the external matrices amounts to the total of
$n(n+1)/2$ real parameters, as required. 

Note that it would be somewhat more elegant
to call the angles in the above factors $\theta_k/2$
instead of $\theta_k$, except the angle $\theta_n$
of the factor ${A}^{sym}_{n,n}$. The reason being
that ${A}^{sym}_{n,n}$ appears only once while    
the others appear twice.

The chain in Eq.(\ref{vnsym}) looks long and perhaps
a bit frightening. However, if needed for practical
applications, it can be somewhat simplified
as we shall now describe.

Consider the case $n=3$, where we introduce
\begin{equation}
V^{(3)sym} \equiv  V^{(2)sym}(\theta_2 / 2)  A^{(3)sym}
V^{(2)sym}(\theta_2 /2)
\end{equation}
where $V^{(2)sym}$ is as defined in Eq.(\ref{v2sym}),
for $n=3$.
Multiplying the factors, we find
\begin{equation}
V^{(3)sym}
= \left(
\begin{array}{ccc} 
c_2-(1-c_3)u^2_1 & is_2 - (1-c_3)u_1u_2 & is_3 u_1 \\
is_2 - (1-c_3)u_1u_2 &  c_2-(1-c_3)u^2_2 & is_3 u_2 \\
is_3 u_1 & is_3 u_2 & c_3
\end{array} \right)
\label{v3sym} 
\end{equation}
Here 
\begin{equation}
u_1 = c^\prime_2 x_1 + i s^\prime_2 x_2, ~~ 
u_2 = c^\prime_2 x_2 + i s^\prime_2 x_1
\label{us}
\end{equation}
$c^\prime_2 = cos(\theta_2/2)$ and 
$s^\prime_2 = sin(\theta_2/2) $.
Moreover, the $x$'s are as introduced in Eq.(\ref{A3sym}).

The essential point is that we may put 
aside the question of the origin of the $u$'s and
their relationship with the $x$'s and simply consider
them as our new variables, as two complex numbers that
satisfy
\begin{equation}
\vert u_1\vert^2 + \vert u_2\vert^2 =1
\end{equation}
Going to the next order, $n=4$,
we may use the identity
\begin{eqnarray}
 A^{sym}_2 A^{sym}_3 A^{sym}_4 A^{sym}_3 A^{sym}_2 &=&
 A^{sym}_2 A^{sym}_3  A^{sym}_2 [( A^{sym}_2)^{-1} 
 A^{sym}_4 ( A^{sym}_2)^{-1}] A^{sym}_2 A^{sym}_3 A^{sym}_2  
\nonumber \\
& = & V^{(3)sym} [A^{\prime sym }_4] 
V^{(3)sym} \nonumber \\ 
A^{\prime sym }_4 & \equiv & ( A^{sym}_2)^{-1} 
 A^{sym}_4 ( A^{sym}_2)^{-1} 
\end{eqnarray}
Here $V^{(3)sym}$ is as found in Eq.(\ref{v3sym}).
From our earlier results, we have
\begin{equation}
A^{sym}_4 = 
\left( \begin{array}{cccc}
1-(1-c_4) y^2_1  & -(1-c_4) y_1 y_2 
& -(1-c_4) y_1 y_3 & is_4 y_1 \\
-(1-c_4) y_1 y_2 & 1-(1-c_4) y^2_2 
& -(1-c_4) y_2 y_3 & is_4 y_2 \\
-(1-c_4) y_1 y_3 & -(1-c_4) y_2 y_3
& 1-(1-c_4) y^2_3  & is_4 y_3  \\
is_4 y_1 & is_4 y_2 & is_4 y_3 & c_4 
\end{array}  \right)
\end{equation}
where $y$'s are real. Therefore, we may 
immediately write down the factor $A^{\prime sym}_4$, 
without having to do any calculations. The first
two components of the vector $y$ get "rotated" but
$y_3$ is untouched. We find
\begin{equation}
A^{sym \prime}_4 = 
\left( \begin{array}{cccc}
c_2-(1-c_4) v^2_1  & -is_2-(1-c_4) v_1 v_2 
& -(1-c_4) v_1 v_3 & is_4 v_1 \\
-is_2-(1-c_4) v_1 v_2 & c_2-(1-c_4) v^2_2 
& -(1-c_4) v_2 v_3 & is_4 v_2 \\
-(1-c_4) v_1 v_3 & -(1-c_4) v_2 v_3
& 1-(1-c_4) v^2_3  & is_4 v_3  \\
is_4 v_1 & is_4 v_2 & is_4 v_3 & c_4 
\end{array}  \right)
\end{equation}
where
\begin{equation}
v_1 = c^\prime_2 y_1 - i s^\prime_2 y_2, ~~ 
v_2 = c^\prime_2 y_2 - i s^\prime_2 y_1, ~~v_3 =y_3
\label{vs}
\end{equation}
Again the vector $v$ has unit norm and we may,
as before, forget about the $y$'s and just use
$v$'s, keeping in mind that $v_1$ and $v_2$
are complex numbers.  
  
As a final example, we wish to compute the quantity $J$
for the case of a three-by-three symmetric matrix,
where
\begin{equation}
(\alpha \beta; jk) \equiv J \sum_{\gamma, i}^{}
\epsilon_{\gamma \alpha \beta} \epsilon_{ijk}
\end{equation}
and $V^{(3)sym}$ is as given in Eq.(\ref{v3sym}).
A glance at this matrix yields
\begin{equation}
J = c_2 c_3 s^2_3 Im (u_2)^2 = c_2 c_3 s_2 s^2_3 x_1 x_2
\label{jsym}
\end{equation}
This resembles our earlier result in Section \ref{simple}
where we found $J^{(3-fam)} = c_2 c_3 s_2  s^2_3 
Im (x^\star_1 x_2)$. The meaning of the $x$'s in the
two cases are, of course, different. Note that a general
$V^{(3)}$ has four parameters while $V^{(3)sym}$
has one less. Equation (\ref{jsym}) is
telling us that the symmetry requirement does not
remove the invariant phase of the matrix. In the
language of Euler rotations, where $V^{(3)}$
would be parameterised with three rotation
angles and one phase, the requirement that the
matrix be symmetric keeps the phase but 
removes one of the rotation angles. Note that the
new phase and angles will be functions of the former
phase and angles.

\end{document}